\documentclass[11pt,a4paper,eps]{article}

\usepackage{graphicx,amssymb}
\usepackage{amsmath}
\usepackage{graphicx}
\usepackage{amsfonts}
\usepackage{amssymb}

\begin{document}

\pagestyle{empty} \addtolength{\topmargin}{20mm}

\begin{center}
{\LARGE Heisenberg Evolution in a Quantum Theory of Noncommutative Fields}
\end{center}

\vskip1.0 cm

\begin{center}
\textbf{Gianluca Mandanici and Antonino Marcian\`{o}} \\[0pt]
\vskip0.3 cm

\textit{Dipartimento di Fisica, Universit\`{a} di Roma ``La Sapienza'', }

\textit{P.le A. Moro 2, 00185 Roma, Italy}
\end{center}

\vspace{1.5cm}

\begin{center}
\textbf{ABSTRACT}
\end{center}

{\leftskip=0.6in \rightskip=0.6in\noindent}

A quantum theory of noncommutative fields was recently proposed by
Carmona, Cortez, Gamboa and Mendez (hep-th/0301248). The
implications of the
noncommutativity of the fields,
intended as the requirements $[\phi,\phi^{+}]=\theta%
\delta^{3}(x-x^{\prime}), [\pi,\pi^{+}]=B\delta^{3}(x-x^{\prime}) $, were
analyzed on the basis of an
analogy with previous results on the so-called ``noncommutative harmonic
oscillator construction".
Some departures from Lorentz symmetry turned out to play a key
role in the emerging framework.
We first consider the same hamiltonian
proposed in hep-th/0301248, and we show that the theory can be
analyzed straightforwardly within the
framework of Heisenberg evolution equation without any need of
making reference to
the ``noncommutative harmonic oscillator construction". We then
consider a
rather general class of alternative hamiltonians, and we observe that
violations of Lorentz invariance are inevitably encountered. These
violations must therefore be viewed as intrinsically associated with
the proposed type of noncommutativity of fields, rather
than as a consequence of a specific choice of Hamiltonian.

\newpage\baselineskip16pt plus .5pt minus .5pt \pagenumbering{arabic} %
\pagestyle{plain}

\addtolength{\topmargin}{-35mm}

\section{Introduction}

Recently a ``quantum theory of noncommutative fields" has been
proposed in Ref.\cite{CCGM}, as a possible field theory
generalization of noncommutative quantum mechanics
\cite{NP,BNS,A}. The type of theory of noncommutative fields
introduced in Ref.\cite{CCGM} is of course rather different from
ordinary relativistic quantum field theory, and, as we will
stress, it is also rather different from a quantum field theory in
noncommutative spacetime. In fact, the theory of Ref.\cite{CCGM}
is formulated in classical (commutative) spacetime, but admits
nonzero equal-time commutation relations not only between the
fields and their conjugated momenta, but also between the field
$\phi$ and its adjoint $\phi^+$, and between the momentum $\pi$
and its adjoint $\pi^+$.

One of the primary reasons of interest in the model proposed in
Ref.\cite{CCGM} resides in its possible use for a phenomenological
description of possible departures from Lorentz symmetry,
potentially relevant for certain classes of observations in
astrophysics~\cite{ABGG}, and of a lack of symmetry between
particles and antiparticles, which could play a role in describing
the observed matter/antimmatter asymmetry.

The theory was analyzed in Ref.\cite{CCGM}, for the case of two
noncommutative fields,  on the basis of an analogy with a
corresponding $2-d$ noncommutative-harmonic-oscillator
problem~\cite{NP,BNS,NH2}. This might suggest that standard
techniques of analysis would not be applicable, and it also
restricts the choice of Hamiltonian to one which is compatible
with the harmonic-oscillator analogy. We intend to show that the
theory can be formulated using the standard techniques based on
the Heisenberg equations of motion. For the case of two fields
with the Hamiltonian adopted in Ref.\cite{CCGM} our formulation
reproduces the results obtained in Ref.\cite{CCGM} on the basis of
the analogy with the $2-d$ noncommutative-harmonic-oscillator
problem. Our formulation however generalizes straightforwardly to
the case of an arbitrary number of fields, and, perhaps more
importantly, provides us the freedom to consider any type of
Hamiltonian.

In the next section we briefly comment on the differences between
the much studied theory of fields in noncommutative spacetimes and
the theory of noncommutative fields proposed in Ref.\cite{CCGM}.
In Section 3, also as a way to introduce some notation, we briefly
review the procedure followed in Ref.\cite{CCGM}, based on the
generalization to field theory of the so-called ``noncommutative
harmonic oscillator construction". In Section 4 we present the
formulation of the theory based on the Heisenberg equation of
motion. In Section 5 we show how the construction can be easily
generalized to the $N$-field case. In Section 6 we consider other
choices of Hamiltonian and comment on the fate of Lorentz symmetry
in this class of theories.

\section{Noncommutative fields vs noncommutative spacetimes}
Since it is not uncommon to
refer to the rather popular field theories in noncommutative spacetimes
as ``noncommutative field theories", we find appropriate to
start our discussion by clarifying the definition proposed
in Ref.\cite{CCGM} for a ``quantum theory of noncommutative fields",
which is basically unrelated to spacetime noncommutativity.

The quantum theory of noncommutative fields~\cite{CCGM} assumes in
addition to the usual quantum, equal time, commutation relations
\begin{align}
[\phi(x),\pi(x')]&=[\phi^+(x),\pi^+(x')]=i\delta^{3}(\vec{x}-\vec{x}^{\prime}),
\label{qc}
\end{align}
also non-vanishing commutators between the fields $\phi$,
$\phi^+$, and between the momenta $\pi$, $\pi^+$:
\begin{align}
[\phi(x),\phi^{+}(x')]&=\theta\delta^{3}(\vec{x}-\vec{x}^{\prime}),  \label{cc} \\
[\pi(x),\pi^{+}(x')]&=B\delta^{3}(\vec{x}-\vec{x}^{\prime}),
\label{cm}
\end{align}
where $\theta$ and $B$ are independent constants. As observed in
Ref.\cite{CCGM}, in order to implement both (\ref{cc}) and
(\ref{cm}) with non-vanishing $\theta$ and $B$ one should resort
to non-hermitian fields.

While in (\ref{cc}) and (\ref{cm}) it is implicitly assumed that
the spacetime coordinates commute (the ``noncommutativity of fields"
is produced by an appropriate expansion of the fields in creation
and annihilation operators), in the better known field theories
in noncommutative spacetime one attributes
to the coordinates noncommutativity that is in general
of the type $[x_{\mu },x_{\nu }]=f_{\mu , \nu}(x)$.
A much studied possibility is the one of a ``canonical
spacetime" (see, {\it e.g.},
Refs.~\cite{susskind,wessliealge,douglasnovikov,gianluKen}),
with coordinate-independent commutator of the
coordinates:
\begin{equation}
\lbrack x_{\mu },x_{\nu }]=i\theta _{\mu ,\nu }.  \label{sc1}
\end{equation}

There is also growing interest in the ``Lie-algebra
spacetimes"~\cite{wessliealge}, with $[x_{\mu },x_{\nu
}]=\gamma^\alpha_{\mu , \nu}x_\alpha$, and particularly in the
$\kappa$-Minkowski
spacetime~\cite{MajidRuegg,lukieAnnPhys,AmelinoMajid,gianlufran,wess2003a}
\begin{align}
\lbrack x_{i},x_{j}]& =0, \\
\lbrack x_{0},x_{i}]& =i\lambda x_{i}.  \label{sc2}
\end{align}

One could ask whether the type of field noncommutativity
(\ref{qc}-\ref{cm}) can be somehow (effectively) introduced
through spacetime noncommutativity  relations of the types
(\ref{sc1}-\ref{sc2}). It easy to verify that this is not the
case. In fact, spacetime noncommutativity immediately leads to the
fact that all the commutators of fields (including the commutator
of a field with itself) are nonvanishing. Consider in particular a
general field expansion of the form:
\begin{equation}
\phi _{a}(x)=\int \frac{d^{4}p}{(2\pi )^{4}}\left( \phi
_{a}(p):e^{ipx}:\right).
\end{equation}%

Choosing, for example, the ordering convention
$:e^{ipx}:=e^{ip_{0}x_{0}}e^{i\vec{p}\vec{x}}$ for exponential
factor, in the case of $\kappa$-Minkowski spacetime one has for
the product of two exponentials:
\begin{align}
\left( :e^{ipx}:\right) \left( :e^{ikx}:\right) & =:e^{i(p\oplus k)x}:, \\
(p\oplus k)& =(p_{0}+k_{0},\vec{p}+\vec{k}e^{-\lambda p_{0}}).
\end{align}%

Thus the commutator of two fields assumes the form
\begin{equation}
\left[ \phi _{a}(x),\phi _{b}(x)\right] =\int \frac{d^{4}p}{(2\pi )^{4}}%
\frac{d^{4}k}{(2\pi )^{4}} \phi _{a}(p)\phi _{b}(k) \left[
e^{i(p_{0}+k_{0})x_{0}}\left( e^{i(\vec{p}+\vec{k}e^{-\lambda p_{0}})\cdot{\vec{x}}}-e^{i(%
\vec{k}+\vec{p}e^{-\lambda k_{0}})\cdot{\vec{x}}}\right) \right],
\end{equation}
which does not vanish and actually depends
on the field ($[\phi,\phi]=f(\phi)$).

An analogous result is obtained starting from the
canonical commutation relations, where one considers
\begin{equation}
\left( :e^{ipx}:\right) \left( :e^{ikx}:\right)
=e^{-\frac{i}{2}p^{\mu }\theta _{\mu ,\nu }k^{\nu }}e^{i(k+p)^{\mu
}x_{\mu }},
\end{equation}%
so that the commutator of two fields reads
\begin{equation}
\left[ \phi _{a}(x),\phi _{b}(x)\right] =\int \frac{d^{4}p}{(2\pi )^{4}}%
\frac{d^{4}k}{(2\pi )^{4}}\phi _{a}(p)\phi _{b}(k)\left[ (-2i)
e^{i(p+k)x} \sin\left(\frac{p^{\mu }\theta _{\mu ,\nu }k^{\nu
}}{2}\right)\right].
\end{equation}

\section{The quantum theory of noncommutative fields from the noncommutative
harmonic oscillator}
The field theory of noncommutative fields proposed in Ref.\cite{CCGM}
was based on the Hamiltonian
\begin{align}
H=\int d\vec{x} \left( \pi^{+} \pi + \vec{\nabla}\phi^{+}\vec{\nabla}%
\phi+m^{2} \phi^{+}\phi\right),
\end{align}
which depends on the fields and on the momenta in
a familiar manner.

We note here that (probably as a result of a typographical error)
the field denoted by $\pi$ in Ref.\cite{CCGM} is not the momentum
conjugate to the field $\phi$. Consistency between the form of the
Hamiltonian and the role of $\pi$ as conjugate momentum is
achieved taking the formulas of Ref.\cite{CCGM} and replacing $\pi
\rightarrow \pi^{+}$, $\pi^{+}\rightarrow \pi$. As one sees from
the form of the commutator (\ref{cm}), this simply leads to a
corresponding change of parameter $B \rightarrow -B$. At the level
of the hamiltonian, the product $\pi^{+} \pi$ is not invariant
under the redefinition $\pi \leftrightarrow \pi^{+}$; however,
since this product appears under spatial integration, the
resulting effect of the above redefinition is the addition of a
constant (-$B$) to the Hamiltonian presented in Ref.\cite{CCGM}.
This constant does not have a physical role, and can therefore be
discarded.

As mentioned, in Ref.\cite{CCGM} the theory was analyzed
rather indirectly, relying on an analogy with the
solution of the quantum-mechanical problem of the noncommutative $2-d$
harmonic oscillator defined by
\begin{align}
H&=\frac{\omega}{2} (q_{1}^{2}+q_{2}^{2}+p_{1}^{2}+p_{2}^{2}), \\
[q_{1},q_{2}]&=i \hat{\theta}, \\
[p_{1},p_{2}]&=i \hat{B}, \\
[q_{i},p_{j}]&=i \delta_{ij},
\end{align}
where $\hat{\theta}=\theta\omega$ and $\hat{B}=B/\omega$.

This noncommutative quantum mechanical system can be transformed
into a corresponding commutative system using the maps
\begin{align}
\frac{q_{1}+i q_{2}}{\sqrt{2}}&=\eta \epsilon_{1} a+ \epsilon_{2} b^{+}, \\
\frac{p_{1}+i p_{2}}{\sqrt{2}}&=-i \epsilon_{1} a+ i \eta \epsilon_{2} b^{+},
\end{align}
where
\begin{align}
\eta &=\sqrt{1+\left(\frac{\hat{B}-\hat{\theta}}{2}\right)^{2}}-\left(\frac{\hat{B}-\hat{\theta}}{2}%
\right), \\
\epsilon_{1}^{2}&=\frac{\hat{B}+\eta}{1+\eta^{2}}, \\
\epsilon_{2}^{2}&=\frac{\eta-\hat{\theta}}{1+\eta^{2}},
\end{align}
with $a,a^{+},b,b^{+}$ satisfying the usual canonical commutation rules
\begin{align}
[a,a^{+}]=[b,b^{+}]=1,
\end{align}
all the other commutators vanishing.

The description of the field $\phi$ and the momentum field $\pi$
was obtained in Ref.\cite{CCGM} assuming an oscillator
of frequency $\omega(p)=\sqrt{\vec{p}^{2}+m^{2}}$
for each value of the momentum $p$:
\begin{align}
\phi&=\int\frac{d\vec{p}}{(2 \pi)^{3}%
\sqrt{2 \omega(p)}}\left[\eta(p) \epsilon_{1}(p) a_{\vec{p}} e^{i \vec{p}%
\cdot\vec{x}}+ \epsilon_{2}(p) b^{+}_{\vec{p}}e^{-i \vec{p}\cdot\vec{x}}%
\right], \\
\pi&=\int\frac{d\vec{p}}{(2 \pi)^{3}}%
\sqrt{\omega (p)} \left[i \epsilon_{1}(p) a^+_{\vec{p}}e^{-i \vec{p}\cdot\vec{x%
}}- i \eta(p) \epsilon_{2}(p) b_{\vec{p}}e^{i \vec{p}\cdot\vec{x}}%
\right].
\end{align}

If $a_{\vec{p}}$ and $b_{\vec{p}}$ satisfy the usual canonical commutation
rules:
\begin{align}
[a_{\vec{p}},a_{\vec{p}^{\prime}}^{+}]=(2\pi)^{3}\delta^{3}(\vec{p}-\vec{p}
^{\prime}), \\
[b_{\vec{p}},b_{\vec{p}^{\prime}}^{+}]=(2\pi)^{3}\delta^{3}(\vec{p}-\vec{p}
^{\prime}),
\end{align}
then the commutation relations (1-3) are satisfied and the
Hamiltonian takes the form:
\begin{align}
H=\int\frac{d\vec{p}}{(2\pi)^{3}}\left[E_{1}(\vec{p})\left( a^{+}_{\vec{p}%
}a_{\vec{p}}+\frac{1}{2}\right)+E_{2}(\vec{p})\left( b^{+}_{\vec{p}}b_{\vec{p%
}}+\frac{1}{2}\right) \right].
\end{align}

This expression for the Hamiltonian indicates that the theory contains free
particles and antiparticles whose energies are respectively
\begin{align}
E_{1}(p)=\omega(p)\left[\sqrt{1+\frac{1}{4}\left(\frac{B}{\omega(p)}-\theta
\omega(p)\right)^{2}}+\frac{1}{2}\left(\frac{B}{\omega(p)}+\theta
\omega(p)\right) \right], \\
E_{2}(p)=\omega(p)\left[\sqrt{1+\frac{1}{4}\left(\frac{B}{\omega(p)}-\theta
\omega(p)\right)^{2}}-\frac{1}{2}\left(\frac{B}{\omega(p)}+\theta
\omega(p)\right) \right].
\end{align}

\section{Quantum theory of noncommutative fields from the Heisenberg
evolution}

In this section we recover the results sketched in the previous section
without resorting to the analogy with the
noncommutative-harmonic-oscillator problem. In addition to
(\ref{qc}), (\ref{cc}) and (\ref{cm}), we will only assume the
validity of the Heisenberg equation of motion:
\begin{align}
\dot{\phi}&=-i[\phi,H],  \label{h1} \\
\dot{\pi}&=-i[\pi,H], \label{h2}
\end{align}
that is necessary upon the identification of the Hamiltonian with the
generator of time evolution.

We observe that by implementing the Heisenberg equation it becomes
legitimate to view $\pi$ as the momentum canonically conjugate to
the field $\phi$. These quantities in fact can be considered as
conjugated only under the equation of motion. We note in
particular that the usual conjugation relation
\begin{equation}
\pi=\dot{\phi}^{+}
\end{equation}
does not hold here.

From equations (\ref{qc}), (\ref{cc}), (\ref{cm}) and
(\ref{h1}-\ref{h2}) one has that:
\begin{align}
\dot{\phi}&=\pi^{+}+i\theta(\vec{\nabla}^{2}-m^{2})\phi,  \label{he1} \\
\dot{\pi}&=(\vec{\nabla}^{2}-m^{2})\phi^{+}-iB\pi.  \label{he2}
\end{align}

Equations (\ref{he1}) and (\ref{he2}) clarify the conjugation relations
between $\phi$ and $\pi$. Substituting (\ref{he2}) in (\ref{he1}) one
obtains the field equation:
\begin{align}
\ddot{\phi}-(1+\theta B)(\vec{\nabla}^{2}-m^{2})\phi-i\{\theta (\vec{\nabla}%
^{2}-m^{2})+B\}\dot{\phi}=0.  \label{mm}
\end{align}
A solution of the above field equation can be obtained with $\phi$ in the
form
\begin{align}
\phi(x) =\int \frac{d\vec{p}}{(2\pi)^{3}} \left[\alpha(p)a_{\vec{p}%
}e^{-i(\omega_{1} t-\vec{p}\cdot\vec{x})}+\beta(p)b^{+}_{\vec{p}}e^{i(\omega_{2}
t-\vec{p}\cdot\vec{x})} \right].  \label{an}
\end{align}

From equation (\ref{he1}) it follows that the conjugate momentum
field must be of the form
\begin{align}
\pi(x) =\int \frac{d\vec{p}}{(2\pi)^{3}} i\left\{\alpha(p)[\omega_{1}-\theta(%
\vec{p}^{2}+m^{2})]a^{+}_{\vec{p}}e^{i(\omega_{1} t-\vec{p}\cdot\vec{x}%
)}-\beta(p)b_{\vec{p}}[\omega_{2}+\theta(\vec{p}^{2}+m^{2})]e^{-i(\omega_{2}
t-\vec{p}\cdot \vec{x})} \right\}.  \label{bn}
\end{align}

The request that the field (\ref{an}) is a solution of the field equation
implies that
\begin{align}
\omega_{1}(p)=\frac{\theta (\vec{p}^{2}+m^{2})-B}{2}\pm \sqrt{\vec{p}%
^{2}+m^{2}+ \left[\frac{\theta (\vec{p}^{2}+m^{2})+B}{2}\right]^{2}}
\end{align}
and that
\begin{align}
\omega_{2}(p)=-\frac{\theta (\vec{p}^{2}+m^{2})-B}{2}\pm \sqrt{\vec{p}%
^{2}+m^{2}+ \left[\frac{\theta (\vec{p}^{2}+m^{2})+B}{2}\right]^{2}}.
\end{align}

Moreover the solution of the field equation (\ref{mm}) must
satisfy the commutation relations (\ref{qc}), (\ref{cc}) and
(\ref{cm}), which imply respectively
\begin{align}
&\alpha^{2}\left[ \omega_{1}-\theta(\vec{p}^{2}+m^{2})\right]+\beta^{2}\left[
\omega_{2}+\theta(\vec{p}^{2}+m^{2})\right]=1, \\
&\alpha^{2}-\beta^{2}=\theta, \\
&\alpha^{2}\left[ \omega_{1}-\theta(\vec{p}^{2}+m^{2})\right]^{2}-\beta^{2}%
\left[ \omega_{2}+\theta(\vec{p}^{2}+m^{2})\right]^{2}=-B.
\end{align}

The first two equations of this system can be solved with respect to $\alpha$
and $\beta$, obtaining
\begin{align}
\alpha^{2}=\frac{1+\theta[\theta(\vec{p}^{2}+m^{2})+\omega_{2}]}{%
(\omega_{1}+\omega_{2})}, \\
\beta^{2}=\frac{1+\theta[\theta(\vec{p}^{2}+m^{2})-\omega_{1}]}{%
(\omega_{1}+\omega_{2})}.
\end{align}

Compatibility with the third equation selects
\begin{align}
\omega_{1}(p)&=\frac{\theta (\vec{p}^{2}+m^{2})-B}{2}+ \sqrt{\vec{p}%
^{2}+m^{2}+ \left[\frac{\theta (\vec{p}^{2}+m^{2})+B}{2}\right]^{2} }, \\
\omega_{2}(p)&=-\frac{\theta (\vec{p}^{2}+m^{2})-B}{2}+ \sqrt{\vec{p}%
^{2}+m^{2}+\left[\frac{\theta (\vec{p}^{2}+m^{2})+B}{2}\right]^{2} }.
\end{align}

It is straightforward now to calculate $\phi$ and $\pi$ using the
above expressions for $\alpha$,$\beta$,$\omega_{1}$,$\omega_{2}$.
The formulas we find exactly reproduce the corresponding ones of
Ref.\cite{CCGM} (briefly discussed in the previous section).

\section{Heisenberg evolution for a theory of noncommutative fields:
the $N$-field case}

In this section we extend the procedure described in the previous
section to the case in which $N$ real fields are involved. First
we must extend the commutation relations (\ref{qc}-\ref{cm}). This
is easily done as follows:
\begin{align}
[\phi_i(x),\phi_j(x')]=i\Theta_{i,j}\delta^{3}(\vec{x}-\vec{x}'), \\
[\pi_i(x),\pi_j(x')]=iB_{i,j}\delta^{3}(\vec{x}-\vec{x}'),
\end{align}
where $i,j= 1, 2, ...N$, and $\Theta_{i,j}, B_{i,j}$ are constant
antisymmetric matrices. Since we are interested in the free
theory, we consider the total Hamiltonian to be the sum of the
Hamiltonian of each field:
\begin{align}
H=\sum_{i=1}^N H_{i}=\sum_{i}\int
d\vec{x}\left(\frac{1}{2}\pi_{i}\pi_{i}+\frac{1 }{2}\vec{\nabla}
\phi_{i}\vec{\nabla} \phi_{i}+\frac{1}{2}m^2\phi_{i}\phi_{i}
\right).
\end{align}

From Heisenberg equation, following the same strategy outlined in
the previous section we obtain the following system of
differential equations:
\begin{equation}
\ddot{\phi}_{i}(x)-\left(\delta_{i,k}+B_{i,j}\Theta_{j,k}\right)\left(
\vec{\nabla}^2-m^2\right)\phi_{k}(x)+\left[
\Theta_{i,j}\left(\vec{\nabla}^2-m^2\right)-B_{i,j}\right]\dot{\phi}_{j}(x)=0.
\end{equation}

This system of equations allows us to observe that,
once the form of the expansions of the fields is taken into account,
a standard Lorentz-invariant form of the energy-momentum
dispersion relations would require vanishing matrices $\Theta_{i,j}$
and $B_{i,j}$, so that the term going like $\dot{\phi}$ would vanish.
But for  $\Theta_{i,j} = B_{i,j} = 0$
the fields of course are no longer ``noncommutative".

\section{Remarks on Lorentz symmetry}

For spacetime noncommutativity there has been much interest in the
emerging departures from classical Lorentz symmetry, signaled by
modified energy/momentum dispersion
relations~\cite{susskind,douglasnovikov,gianluKen,AmelinoMajid,gianlufran,dsr}.
The presence of departures from classical Lorentz covariance of
the ``theory of noncommutative fields" here considered, was
already pointed out in Ref.\cite{CCGM} (see also
Ref.\cite{CCGMLV}), indeed through an analysis of the
energy/momentum dispersion relations. In our approach the
violation of Lorentz symmetry is evident at the level of the field
equation (\ref{mm}). One could ask if the violation of Lorentz
covariance is necessarily implied by the switching-on of the
nontrivial $[\phi,\phi^{+}]$ (and $[\pi,\pi^{+}]$) commutators. In
particular, one might wonder whether an appropriate choice of
Hamiltonian might compensate for the $\theta$, and $B$,
Lorentz-violating factors in the fields equations coming from the
commutation rules.

We consider the rather general class of Hamiltonians of the form:
\begin{align}
H=\int d\vec{x}  & \left\{ \alpha_{1}(\theta,B,m) \pi \pi^{+}+
\alpha_{2}(\theta,B,m) \phi \phi^{+}+ \alpha_{3}(\theta,B,m)
\vec{\nabla} \phi \vec{\nabla} \phi^{+}+ \alpha_{4}(\theta,B,m)
(\pi \phi
+\phi^{+}\pi^{+})+\right. \nonumber \\
&\left.+\alpha_{5}(\theta,B,m)(\pi \phi^+ + \phi \pi^+)+
\alpha_{6}(\theta,B,m) \vec{\nabla} \pi \vec{\nabla} \pi^{+} +
\alpha_{7} (\theta,B,m) (\vec{\nabla}^{2} \pi \phi+\phi^{+}
\vec{\nabla}^{2}\pi^{+})+ \right. \nonumber \\
&\left.+\alpha_{8}(\theta,B,m)(\vec{\nabla}^{2} \pi^+ \phi
+\phi^+\vec{\nabla}^{2}\pi) \right\}.
\end{align}

The Heisenberg equations then read
\begin{align}
&\dot{\phi}=\mathbf{A}_1 \phi +\mathbf{B}_1 \phi^+ + \mathbf{C}_1 \pi + \mathbf{D}_1 \pi^+  , \label{eqev1} \\
&\dot{\pi}=\mathbf{A}_2 \phi +\mathbf{B}_2 \phi^+ + \mathbf{C}_2
\pi + \mathbf{D}_2 \pi^+ , \label{eqev2}
\end{align}
where
\begin{align}
\mathbf{A}_1&=\alpha_{4}-i \alpha_{2} \theta + (\alpha_{7}+i \alpha_{3}\theta)\vec{\nabla}^{2},  \label{A1} \\
\mathbf{B}_1&=\alpha_{5}+\alpha_8 \vec{\nabla}^{2}, \\
\mathbf{C}_1&=-i\theta \alpha_{5}-i\theta\alpha_8 \vec{\nabla}^{2}, \\
\mathbf{D}_1&=\alpha_{1}-i \alpha_{4} \theta -(\alpha_{6}+i \alpha_{7}\theta)\vec{\nabla}^{2},  \label{B1} \\
\mathbf{A}_2&=-i B \alpha_{5}-i B \alpha_8 \vec{\nabla}^{2},\\
\mathbf{B}_2&=-\alpha_{2}-i \alpha_{4} B +(\alpha_{3}-i \alpha_{7}B)\vec{\nabla}^{2},  \label{C1} \\
\mathbf{C}_2&=-\alpha_{4}-i \alpha_{1} B -(\alpha_{7}-i
\alpha_{6}B)\vec{\nabla}^{2},  \label{D} \\
\mathbf{D}_2&=- \alpha_{5}- \alpha_8 \vec{\nabla}^{2}.
\end{align}

We start by analyzing the simple case of $\alpha_5=\alpha_8=0$. In
this case from (\ref{eqev1})-(\ref{eqev2}) one obtains the field
equations:
\begin{align}
\ddot{\phi}=[ \mathbf{D}_1 \mathbf{B}_2^{+}-
\mathbf{A}_1 \mathbf{C}^{+}_2 ] \phi + [\mathbf{A}_1+\mathbf{C}_2^{+}] \dot{\phi}, \label{feg1} \\
\ddot{\pi}=[\mathbf{D}_1 \mathbf{B}_2^{+}-\mathbf{A}_1
\mathbf{C}^{+}_2 ] \pi + [\mathbf{A}_1+\mathbf{C}_2^{+}]
\dot{\pi}. \label{feg2}
\end{align}

A necessary condition for the covariance of the above equations is
that
\begin{align}
\mathbf{A}_1=-\mathbf{C}_2^{+},
\end{align}
which implies the relations
\begin{align}
\alpha_{2} \theta= \alpha_1 B  \\
\alpha_{3} \theta= \alpha_6 B.
\end{align}

These relations, for nonzero $\theta$ and $B$, are not compatible
with the proper $\theta,B \rightarrow 0$ limit
($\alpha_{1}=\alpha_{3}=1$, $\alpha_2=m^2$,
$\alpha_4=\alpha_6=\alpha_7=0$).

In the case $\alpha_5,\alpha_8 \neq0$ the field equations assume a
much more involved form than that of (\ref{feg1}) and
(\ref{feg2}), but again it can be shown that, for nonzero $\theta$
and $B$, the field equations are incompatible with Lorentz
covariance. In order to achieve less severe departures from
Lorentz symmetry, while still working within the framework of
Heisenberg evolution, one might have to adopt field commutation
relations different from the ones here considered (see e.g.
Ref.\cite{HEDCR}).

\section*{Acknowledgements}
The authors thank Giovanni Amelino-Camelia for bringing the problem
to their attention and for valuable discussions.

\end{document}